\documentstyle[preprint,aps]{revtex}
\begin{document}
\draft


\widetext
\title{ Structure and stability of bosonic clouds: \\
alkali atoms with negative scattering length
}
\author {A. Parola$^{1,2}$, L. Salasnich$^{1,3}$, L. Reatto$^{1,3}$}
\address{
$^1$ Istituto Nazionale per la Fisica della Materia, Unit\`a di Milano, 
Via Celoria 16, Milano, Italy\\
$^2$ Istituto di Scienze Fisiche, Universit\'a di Milano, Via Lucini 3 Como,
Italy \\
$^{3}$  Dipartimento di Fisica, Universit\'a di Milano,
Via Celoria 16, 20133, Milano, Italy }
\maketitle
\begin{abstract}
We investigate the form and stability of a cloud of atoms 
confined in a harmonic trap when the scattering length is negative.
We find that, besides the known low density metastable solution,
a new branch of Bose condensate appears at higher density when
non locality effects in the attractive part are  taken into account. 
The transition between the two classes of solutions as a function
of the number $N$ of atoms can be either sharp or smooth according to 
the strength and range of the attractive interaction. Use of tight 
traps is favorable for investigating the evolution of the system
as the strength of the effective interaction increases with $N$.
\end{abstract}
\narrowtext
Few alkali isotopes are characterized by a negative scattering length 
$a_T$ at low energy in the triplet channel \cite{nega}. 
This is believed to have
important consequences on the existence and the structure of the
Bose Einstein condensed (BEC) state for trapped clouds of atoms when
the isotope follows Bose statistics. The most celebrated case is
$^7$Li for which there is experimental evidence in favor of a BEC state
if the number of atoms does not exceed a threshold \cite{bradley}.
This behavior was in fact predicted theoretically. When the interaction
between atoms is represented by an effective local potential
$v_{\rm eff}({\bf r})=(4\pi\hbar^2/m)a_T\delta^3({\bf r})$, 
a uniform gas phase is unstable towards 
the formation of collapsed state of high density.
The situation changes in presence of an external trap \cite{burnett}. 
Within the Gross-Pitaevskii (GP) equation 
for the condensate wavefunction \cite{gross}, the kinetic energy due to the 
confinement in the trap opposes 
the collapse as long as the number $N$ of atoms does not exceed a 
critical number $N_c$ which depends on the parameters of the system
(isotope mass $m$, scattering length $a_T$, shape of the trap, etc.) 
\cite{baym}. 
Even for $N< N_c$, we must recall that a dilute cloud 
represents a metastable state which can be experimentally observed
only if its lifetime is long enough. 

The possibility of having positive as well as negative scattering lengths 
in different atomic systems arises because the potential between a pair 
of atoms generally supports bound states leading to the formation of
a dimer with several vibrational levels.
Under this condition, $|a_T|$ can be very large but at the same time the 
energy dependent scattering cross section $\sigma_T(E)$
starts to deviate from its zero energy limit $\sigma_T(0)=4\pi a_T^2$
already at an energy that can be extremely small \cite{australi}. 
The calculations by C\^ot\'e {\it et al.}
\cite{dalgarno} for $^7$Li show that $\sigma_T(E)$ is reduced to 50\%
of $\sigma_T(0)$ at energies of the order of $3\,10^{-9}$ a.u. . This value is
small enough that $s$-wave scattering is dominant over higher angular 
momentum states meaning that the scattering process between two atoms can 
still be represented by an effective two body interaction $v_{\rm eff}$.
However, in general we cannot neglect the momentum dependence of
$v_{\rm eff}(k)$ for the colliding atoms. 
This amounts to say that the effective interaction is non local:
In this note we investigate the effects of non-locality on the stability of
the cloud.

It is rather easy to convince that non-locality in the interatomic
effective potential can modify the stability condition of the Bose
condensate. The GP functional ${\cal E } [\Psi]$ with non local 
interaction $v_{\rm eff}({\bf r})$ reads:
\begin{eqnarray}
{\cal E}[\Psi] = 
&&\int d^3 {\bf r} {\hbar^2\over 2 m} \vert \nabla \Psi\vert^2
+\int d^3 {\bf r} U_{\rm ext}({\bf r}) \,
\vert\Psi({\bf r})\vert^2\, +\nonumber\\
&&{1\over 2} \int d^3 {\bf r} \int d^3 {\bf r}^\prime 
\vert\Psi({\bf r}^\prime)\vert^2\,
v_{\rm eff}({\bf r}-{\bf r}^\prime)\, \vert\Psi({\bf r}) \vert^2 
\label{gpf}
\end{eqnarray}
where $\Psi({\bf r})$ is the wavefunction of the condensate and 
$U_{\rm ext}({\bf r})$ is the potential of the trap. In this paper 
we consider a symmetric harmonic trap $U_{\rm ext}({\bf r})=
{1\over 2}m\omega_0^2 r^2$. In the local limit, 
one recovers the standard form 
of the GP functional. The ground state wavefunction of a cloud of $N$ atoms
is determined  by minimizing ${\cal E}[\Psi]$ with the constraint that 
\begin{equation}
\int d^3 {\bf r} \vert\Psi({\bf r})\vert^2\,=N 
\label{norm}
\end{equation}
In the ground state $\Psi({\bf r})$ is positive definite 
and spherically symmetric.

As a first step, we discuss an approximate, variational approach to
this problem which already shows the main features of the exact solution.
As a trial wavefunction we choose a Gaussian form:
\begin{equation}
\Psi({\bf r})=N^{1\over 2}\,
\left ({\lambda^2 \over\pi a_H^2}\right )^{3\over 4}
\exp\left (-{\lambda^2 r^2\over 2 a_H^2}\right ) 
\label{gauss}
\end{equation}
with a single variational parameter $\lambda$ which defines the size of
the cloud in units of the harmonic oscillator length 
$a_H=[\hbar/(m\omega_0)]^{1/2}$. By substituting 
${\bf r}^\prime={\bf r}+{\bf s}$ in Eq. (\ref{gpf}) and developing
$\Psi({\bf r} +{\bf s})$ in powers of ${\bf s}$, we perform a gradient
expansion of the energy functional ${\cal E}[\Psi]$. 
To lowest order in the gradient of the wavefunction, the energy reads:
\begin{equation}
{\cal E}(\lambda ) ={\hbar\omega_0 \over 2}N \left ( {3\over 2}\lambda^2 +
{3\over 2}\lambda^{-2} + \gamma\lambda^3 +\gamma\,\tau \lambda ^5 \right ) 
\label{vare}
\end{equation}
where the two parameters $\gamma$ and $\tau$ depend on the choice of the
effective interaction providing a dimensionless estimate of the
strength $\gamma$ and range $\tau$ of $v_{\rm eff}(r)$:
\begin{eqnarray}
\gamma&=&\sqrt{2/\pi} N \,a_T\,a_H^{-1} \nonumber\\
\tau&=&-\left [3\int dr r^4 v_{\rm eff}(r)\right ]
\,\,\left [2a_H^2 \int dr r^2 v_{\rm eff}(r)\right ]^{-1}
\end{eqnarray}
According to these definitions, $\tau$ is basically minus the 
square of the range of the effective interaction and is therefore
a negative quantity. For a local interaction $\tau=0$ and we recover 
Baym and Pethick result \cite{baym}. 
In this limit and for negative scattering length, $\gamma <0$ and 
${\cal E}(\lambda)$ goes to $-\infty$ for $\lambda\to\infty$ leading
to a collapsed ground state. However, ${\cal E}(\lambda)$ has a local
minimum for a finite value of $\lambda$ when $|\gamma |< 4/5^{5/4}$, i.e.
when $N< N_c\sim 0.67 \,\,a_H \,|a_T|^{-1}$. 
This represents the metastable state mentioned above.

It is clear from Eq. (\ref{vare}) that non-locality changes the stability 
condition because, for negative scattering length,
the product $\gamma\tau$ is positive.
Depending on the value of the parameters, ${\cal E}(\lambda)$ can have one or
two minima  and the collapse is prevented in any case. However, this result is
only suggestive because the gradient expansion breaks down as soon as 
non-locality becomes important and a more appropriate formalism is needed.
We have then studied the variational equation obtained
with the same choice of a Gaussian 
wavefunction (\ref{gauss}) but without invoking gradient expansion.
This problem requires the explicit definition of the effective interaction.
We assume that the attractive potential has a finite range $r_e$ and
in addition we allow for the presence of a repulsive contribution
which is modeled as a {\it local} positive term defined by a ``high energy"
scattering length 
$a_R > 0$. With this choice, $v_{\rm eff}(k)$ changes sign from negative at
small $k$ (i.e. low energy), to positive at larger $k$ thereby mimicking
the microscopic computations of the $\ell=0$ phase shift $\delta_k$ 
which show a change in sign from positive to negative when $k$ increases
\cite{australi}.
At still higher momenta, scattering in other channels with $\ell\ne 0$
becomes important but we assume that the range of density is such that 
this regime is never reached. 
\par
In the model we study, the effective interaction is then written 
in the following form:
\begin{equation}
v_{\rm eff}(k)={4\pi\hbar^2\over m}\left [a_R + (a_T-a_R)\,f(kr_e) \right ]
\label{effect}
\end{equation}
We have considered two choices for the shape function $f(x)$:
A Lorenzian $f(x)=(1+x^2)^{-1}$ and a Gaussian $f(x)=\exp(-x^2)$.
The results do not depend on the specific choice of $f(x)$ and so we 
will discuss only the Lorenzian case. 
We use interaction parameters appropriate for $^7$Li: $a_T=-27\,a_B$ 
\cite{abraham}, $r_e=10^3\,a_B$ \cite{dalgarno} and $a_R=6.6\,a_B$
\cite{note} (where $a_B$ is the Bohr radius). The energy functional for the 
Lorenzian potential can be analytically expressed in terms of elementary
functions and reads:
\begin{eqnarray}
{\cal E}(\lambda )=&&{\hbar\omega_0 \over 2}N \Big\{ {3\over 2}\lambda^2 +
{3\over 2}\lambda^{-2} + \gamma_R\lambda^3 - 
\tau_1 \lambda + \nonumber \\
&&+\tau_2 {\rm erfc}\left (\chi\lambda^{-1}\right )
\exp\left(\chi^2\lambda^{-2}\right ) \Big\} 
\label{vare1}
\end{eqnarray}
where 
\begin{eqnarray}
\gamma_R&=&\,N\,\sqrt{2\over\pi}\,{a_R\over a_H}\qquad
\tau_1=N\,\sqrt{2\over\pi}\,a_H\,{a_T-a_R\over r_e^2} \nonumber \\
\chi&=&{a_H\over\sqrt{2}r_e} \qquad\qquad\quad
\tau_2=N\,a_H^2\,{a_T-a_R\over r_e^3} \nonumber
\end{eqnarray}
and erfc$(x)=1-{\rm erf}(x)$ is the complementary
error function. The extrema of ${\cal E}(\lambda)$ 
are obtained as solutions of an algebraic equation. This equation has 
either one or three positive roots depending on the parameters and
on number $N$ of atoms in the cluster. 
When three solutions are present, the 
intermediate one represents an unstable state (i.e. a local maximum of the 
energy) while the other two respectively describe a low density metastable 
solution and a minimum which represents the stable solution within 
GP approximation.

The variational results for three typical trap
sizes are shown in Fig. 1 where the average radius and the density at 
the center of the cloud are
plotted as a function of $N$. In the same figure,
the variational data are  also compared with the exact solution of the 
GP equation, obtained by numerical 
integration of the corresponding self-consistent Schr\"odinger equation. 
In fact, the exact minimization of the GP functional (\ref{gpf}) gives
rise to a non linear eigenvalue problem for the ground state
wavefunction. The ground state is spherically symmetric and can
be written as $\Psi ({\bf r})=\phi (r)/r$ where $\phi(r)$ satisfies:
\begin{equation}
-{\hbar^2\over 2 m}{{\rm d}^2\over {\rm d}r^2}
\phi(r)  + {m\omega_0^2 \over 2} r^2 \phi (r) 
+ W(r) \phi (r) = \epsilon \phi (r) 
\label{nls}
\end{equation}
Here $\epsilon$ is the chemical potential and 
\begin{equation}
W(r)= \int d^3{\bf r}' \; |\phi (r')|^2 v_{\rm eff}({\bf r}-{\bf r}') 
\end{equation}
is the self-consistent interaction. For effective potentials written 
as sums of Yukawa functions, this integro-differential equation can 
be reduced to a boundary value problem for a set of ordinary differential 
equations. In particular, in the case of interest, the form (\ref{effect}) 
gives rise to three ordinary differential equations which have been
integrated by standard Runge-Kutta method. 

Fig. 1 shows the good performance of the variational approach which 
is always very close to the exact solution.
For large $N$ the size of the cloud is remarkably 
independent of the trap size demonstrating that the atoms 
are in a self-bound configuration. The size of the cloud is governed by
the effective range: For large values of $r_e$ ($r_e\gg |a_T|$) and
$N$ not too large, the reduced density $\rho |a_T|^3$ is small. This 
indicates that the self bound state that we find represents a novel regime
of the cloud, intermediate between the very low density state already 
predicted within the approximation of local interaction and the collapsed
high density state which depends on the detailed shape of the true 
interatomic potential \cite{stoof}.

The asymptotic large $N$ 
behavior of the exact solution can be obtained analytically
from our equation: The radius of the cloud reaches a finite limit 
which coincides with the result of the Thomas-Fermi approximation to the 
GP equation \cite{lewe}. On the other hand, the
low density branch is virtually indistinguishable from the 
exact solution of a local attractive potential, also shown in figure.
The effects of non-locality become important just when the radius of 
the cloud rapidly drops. This ``transition" is discontinuous for 
large traps, where the reentrant behavior of the curve shows the
presence of an unstable branch. By reducing the trap size, however, this
discontinuity is strongly reduced and, below $0.3\,\mu$m, the unstable
branch disappears and there is a smooth evolution from a very
dilute cloud to a less dilute state with an increasing density as
$N$ grows. 

The predictions of the standard treatment of clouds of alkali atoms
in terms of a local pseudopotential have to be modified in two respects when
the scattering length is negative. In the case of shallow traps, the
stability threshold $N_c$ is only an upper bound. There is a lower
threshold $N_{cl}$ for the higher density state and in the intermediate 
region $N_{cl} < N < N_c$ the low density branch is metastable towards 
the higher density solution. For instance, in a $3\,\mu$m trap 
$N_c=1300$ and $N_{cl}=160$. Which state is reached experimentally 
depends on how the cloud is formed. In the case of a
tight trap a threshold does not exist altogether. 
This feature allows to explore
experimentally how the state of the system evolves as the interaction 
strength increases and depletion effects start to set up even at the 
lowest temperature.
It should be noticed that the presence of a repulsive short range 
interaction in our model potential does not play a crucial role.
The higher density branch as well as the existence of two regimes depending on
the size of the trap are present even if we put $a_R=0$. The only difference
is that, in this case, the radius of the cloud slowly decreases as 
$N\to \infty$ and does not reach a finite limit as in Fig. 1. In any case, 
it should be kept in mind that in the large $N$ limit our results are
only qualitative because the GP equation itself breaks down and
interaction effects are expected to produce a depletion of the 
condensate when $\rho |a_T|^3$ is not very small. 
However, we believe that the trend shown by this
equation is significant even if quantitative results must wait for
a more accurate determination of the interatomic potential and 
a better treatment of of the many body problem.

Another interesting problem which can be addressed by this formalism is the
shape of the condensate wavefunction. This information may be useful
for the quantitative determination of the number of atoms in the condensate
from experimental data. The GP ground state wavefunction, normalized
to unity at the origin, is shown for different numbers of atoms in the cloud
in Fig. 2a. We see that some deviation from a Gaussian form
develops as density grows. In particular, a sharp peak at the cloud center
grows at large $N$. Along the very low density branch we verified
that the wavefunction is not affected by the non locality of the 
interaction and is very close to the numerical result in the local
approximation. However, significant deviations appear as soon as
we approach the edge of the low density branch. 

As a next step we can investigate the dependence of the shape of 
the equilibrium curve when the range of the attractive interaction
is varied. In fact, we expect that the larger is this lengthscale,
the stronger is the effect of non locality 
in the stability diagram of the cloud. 
More importantly, the value of the effective range $r_e$ is rather 
uncertain even in Lithium and it is useful to test the sensitivity 
of our results to variations of the shape of the effective interaction.
In Fig. 2b we show the equilibrium curves of a cloud of Lithium atoms in a
harmonic trap of $a_H=1\,\mu$m when the effective range $r_e$
is artificially halved or doubled with respect to the reference
value $r_e=10^3\,a_B$ previously adopted. The results clearly 
show a smoothing of the transition between the two branches 
induced by the enhancement of non locality. By increasing the
effective range, the maximum number of atoms in the low density branch
$N_c$ slightly increases and the two branches of solutions tend to merge
into a unique phase smoothly connecting the low and the higher
density regimes. A quantitative analysis of the variational 
solution shows that the occurrence of the single branch regime
is mainly determined by the ratio between the trap radius and the 
effective range, independently of the scattering length: $a_H < 5\,r_e$.
Conversely, the number of atoms in the cloud at the inflection point
can be estimated as $N_c\sim 4.5 \, r_e\,|a_T|^{-1}$, independently of the
trap radius. These results show that the effective range plays a 
crucial role in determining the stability of the bosonic cloud 
at intermediate density. The previous estimates will become particularly
valuable
when an accurate determination of the interaction parameters will become
available also for other alkali isotopes with negative scattering 
length. Interestingly, new techniques are being devised in order to 
modify the effective interaction in these systems \cite{kagan}.

In conclusion, the analysis of the Bose-Einstein condensed ground state 
of Lithium atoms in a harmonic trap has shown that 
atomic clouds may exist in a novel state, intermediate in density between
the very dilute regime which has been studied up to now and the collapsed
high density state which is determined by the short range behavior 
of the true interatomic potential. Also this intermediate state is
self-bound because the size of the cloud is essentially independent
of the size of the trap. The importance of the study of this intermediate
regime arises from the possibility of tuning the strength of the interaction 
by changing $N$: In this way it will be possible to study how the 
depletion of the condensate affects the structure and the physical properties 
of the state. To this end, it is important to
confine the system in sufficiently small traps in order to 
avoid the discontinuous jump between the two phases.
Interestingly, such micro-traps are under development \cite{haensch}.
Non locality in the effective interaction is a crucial 
ingredient which prevents collapse when the scattering length
is negative. 
Long range potentials favor the formation of big clouds which 
still maintain a rather low density. Therefore, it is quite interesting 
to select elements and isotopes characterized by negative
scattering length together with long range effective interactions.
We stress that our results 
apply to the dilute regimes where GP equation correctly represents
the behavior of the ground state. At higher density, collisions between
atoms deplete the condensate. In addition three body scattering starts to
play a role and spin-flip processes lead to rapid destruction of the 
cloud.

We acknowledge financial support from the ``BEC" advanced research project of 
INFM.

\vfill\eject
\centerline{\bf Figure captions}

\par\noindent
{\bf Fig. 1}
Average radius of the cloud $R$ (panels a,c,e) and central density (panels 
b,d,f) as a function of number of atoms $N$
for the trap sizes $a_H=3\,\mu$m, $1\,\mu$m and $0.3\,\mu$m respectively. 
Full dots represent the numerical solution 
of the GP equation with non local Yukawa potential. Open dots 
correspond to the local approximation. Solid line is the variational result
and the dashed line is the asymptotic limiting radius for $N\to\infty$.
Parameters have been chosen as to mimic $^7$Li.

\vskip 0.5 true cm
\par\noindent
{\bf Fig. 2} Panel (a): Condensate wavefunction $\Psi(r)$ 
for a trap size $a_H=0.3\,\mu$m
and non local interaction. The non interacting result (Gaussian 
wavefunction) would be a straight line in this plot. 
Dashed line: $N=68$ (low density). 
Dotted line: $N=159$ (near the maximum of the low density branch). 
Solid line: $N=166$ (higher density branch).
Panel (b): Equilibrium curves for parameters appropriate for Lithium atoms 
in a trap of $a_H=1\,\mu$m and 
three choices of the effective range of the Lorenzian attractive
potential. Variational results.
Dotted line: $r_e=500\,a_B$. Solid line $r_e=1000\,a_B$.
Dashed line $r_e=2000\,a_B$.

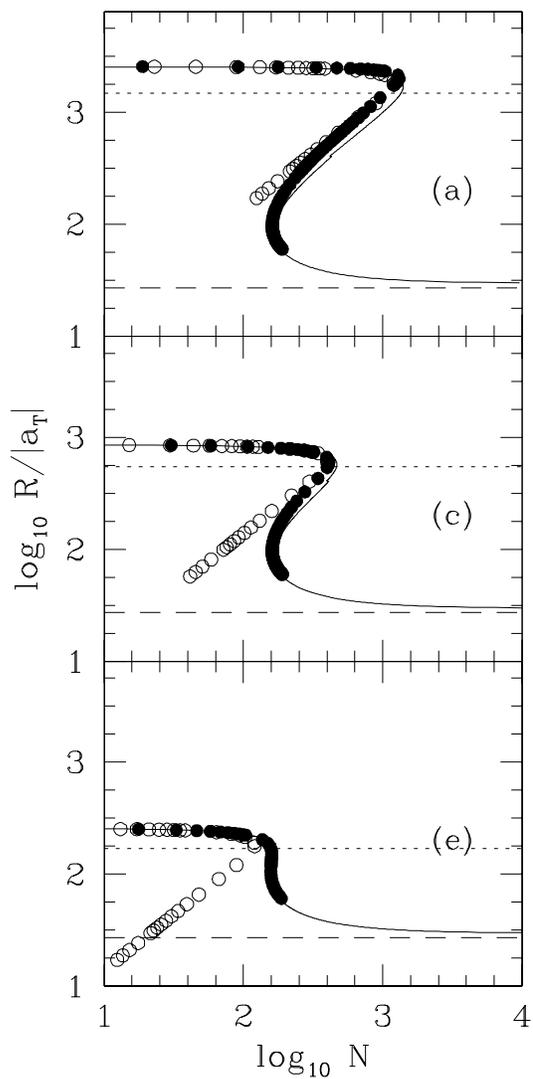
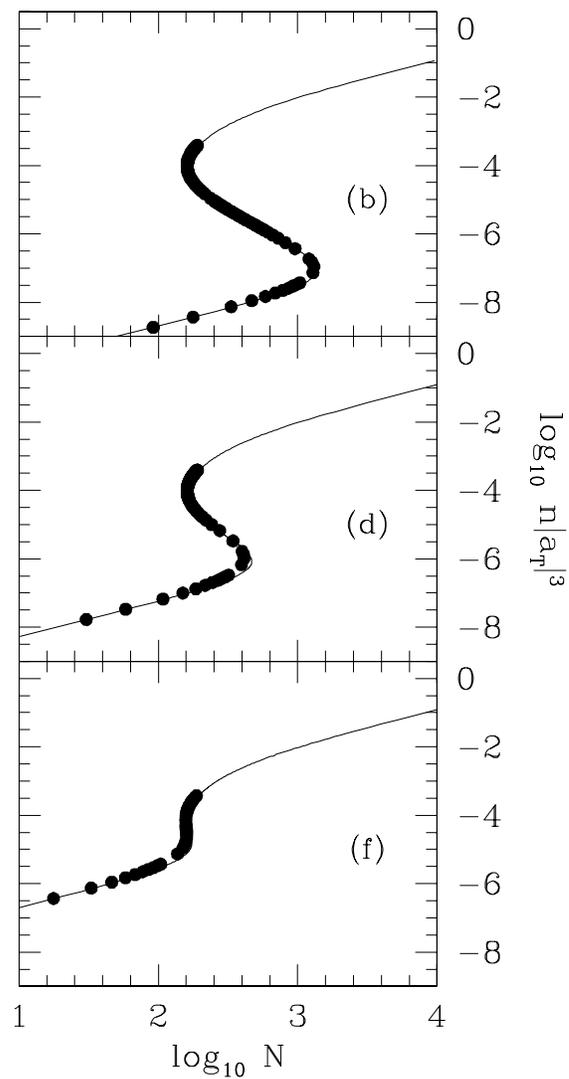
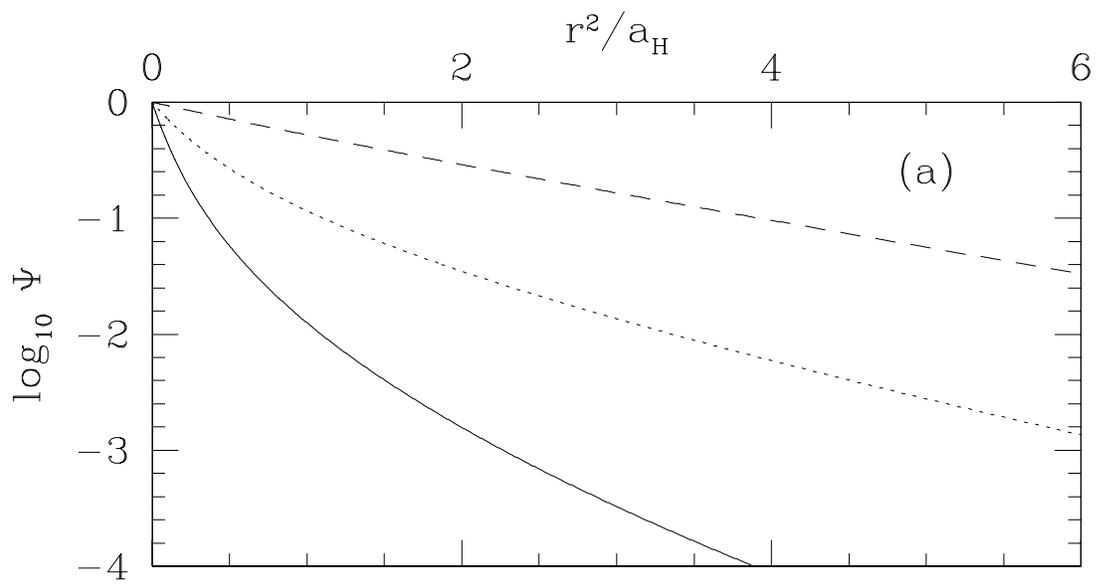
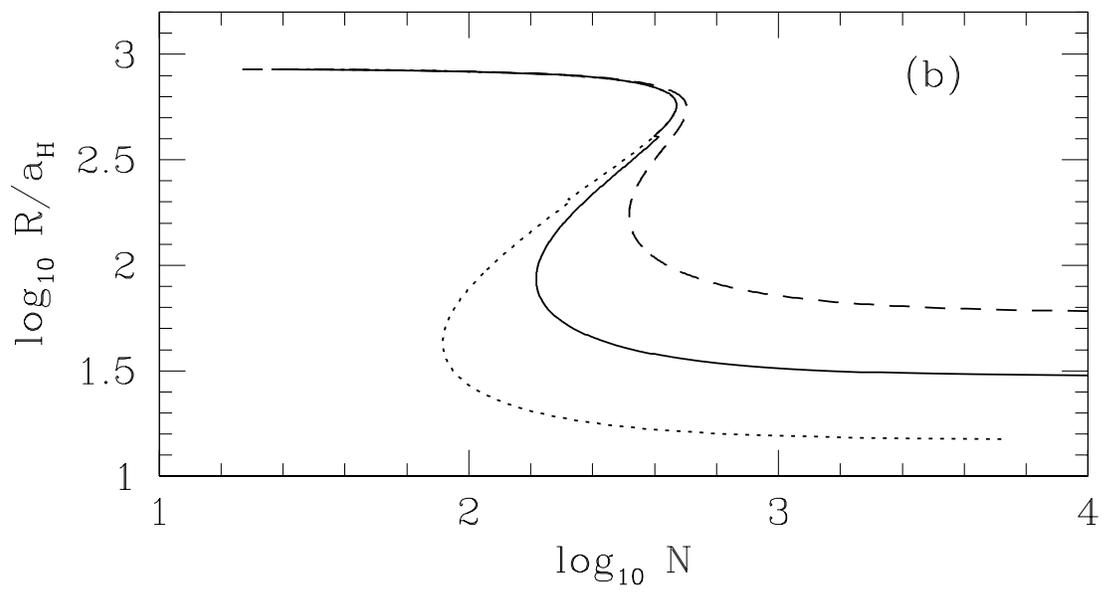
\end{document}